\documentclass[preprint,showpacs,preprintnumbers,amsmath,amssymb,natbib]{revtex4}
\usepackage{graphicx}
\usepackage{dcolumn}
\usepackage{bm}
\usepackage[T1]{fontenc}
\def\bb#1{\mbox{\footnotesize $(#1)$}}

\begin{document}

\title{Maximal correlation between flavor entanglement and oscillation damping due to localization effects}

\author{V. A. S. V. Bittencourt}
\email{vbittencourt@df.ufscar.br}
\author{C. J. Villas Boas}
\email{villasboas@ufscar.br}
\author{A. E. Bernardini}
\email{alexeb@ufscar.br}
\affiliation{Departamento de F\'{\i}sica, Universidade Federal de S\~ao Carlos, PO Box 676, 13565-905, S\~ao Carlos, SP, Brasil}

\begin{abstract}
Localization effects and quantum decoherence driven by the mass-eigenstate wave packet propagation are shown to support a statistical correlation between quantum entanglement and damped oscillations in the scenario of three-flavor quantum mixing for neutrinos.
Once the mass-eigenstates that support flavor oscillations are identified as three-{\em qubit} modes, a decoherence scale can be extracted from correlation quantifiers, namely the entanglement of formation and the logarithmic negativity.
Such a decoherence scale is compared with the coherence length of damped oscillations.
Damping signatures exhibited by flavor transition probabilities as an effective averaging of the oscillating terms are then explained as owing to loss of entanglement between mass modes involved in the relativistic propagation.
\end{abstract}

\date{\today}
\maketitle

\paragraph{Introduction.}

The quantum oscillation framework \cite{01b01} driven by mixing properties \cite{01b02, 01b03} supports, almost sufficiently, the enlarged phenomenological scenario of flavor oscillations in neutrino physics.
The flavor transition between observable quantum states involved in the production and detection of oscillating neutrinos can be accomplished by decoherence mechanisms which are effectively evinced, for instance, in a simplified external wave packet framework \cite{01b01,01b05,01b06}.
Localization effects indeed lead to the spatial quantum decoherence that imprints damping signatures onto neutrino oscillation probabilities \cite{Ber003,Ber004}.

These damping signatures are commonly attributed to some small corrections due to non-oscillation effects \cite{Ber005} that result from quantum decoherence, quantum decay, or even exotic oscillation mechanisms, and they are classified according to their spectral dependencies \cite{01b07}.

One might suppose that a complete physical understanding of the flavor oscillation damping emerges when one changes the view of single particle quantum mechanics to that one of composite quantum system framework \cite{01b08}.
It provides an efficient theoretical tool through which damping signatures and related features can be extracted \cite{Ber13,01b13}.
Our purpose in this Letter is to identify the statistical correlation between the coherence length related to the flavor oscillation damping and the decoherence scale supported by entangled mass modes that drive flavor oscillations into a composite quantum system framework.

Entanglement is a natural quantum correlation that arises as consequence of the superposition principle in composite quantum systems \cite{01b09,01b10}.
In particle physics, quantum entanglement has already been considered for quantifying the particle mixing in two-body systems like $K_0 \, \overline{K}_0$ and $B_0 \, \overline{B}_0$ states produced in electron-positron annihilations \cite{01b11,01b12}.
Recently, it has been shown that a related framework can be extended to the domain of quantum field theories \cite{Nishi,Blasone01,BerFor}, where a fine structure of quantum correlations associated to multi-mode and multi-particle entanglement has been identified \cite{Blasone3}.
Hereafter entanglement of formation ($E_{F}$) and logarithmic negativity ($E_{\mathcal{N}}$) are reported as quantifiers for quantum correlations between mass fields in the scenario of three-flavor neutrino mixing.

Neutrino mass-eigenstates are then supposed to compound a $3$-dim orthonormal basis as to make possible the description of each mass-eigenstate as a three-{\em qubit} state \cite{01b10}.
It supports the picture of flavor states described by entangled states such that the above mentioned quantum correlations can be computed between mass modes \cite{01b10}.
It allows one to obtain a statistical correlation between entanglement and damping as a function of a localization parameter, $\sigma_p$, which parameterizes the wave packet width as $\propto 1/\sigma_p$.

\paragraph{Flavor oscillations and localization effects.}

\begin{figure}
\includegraphics[width = 7.0 cm]{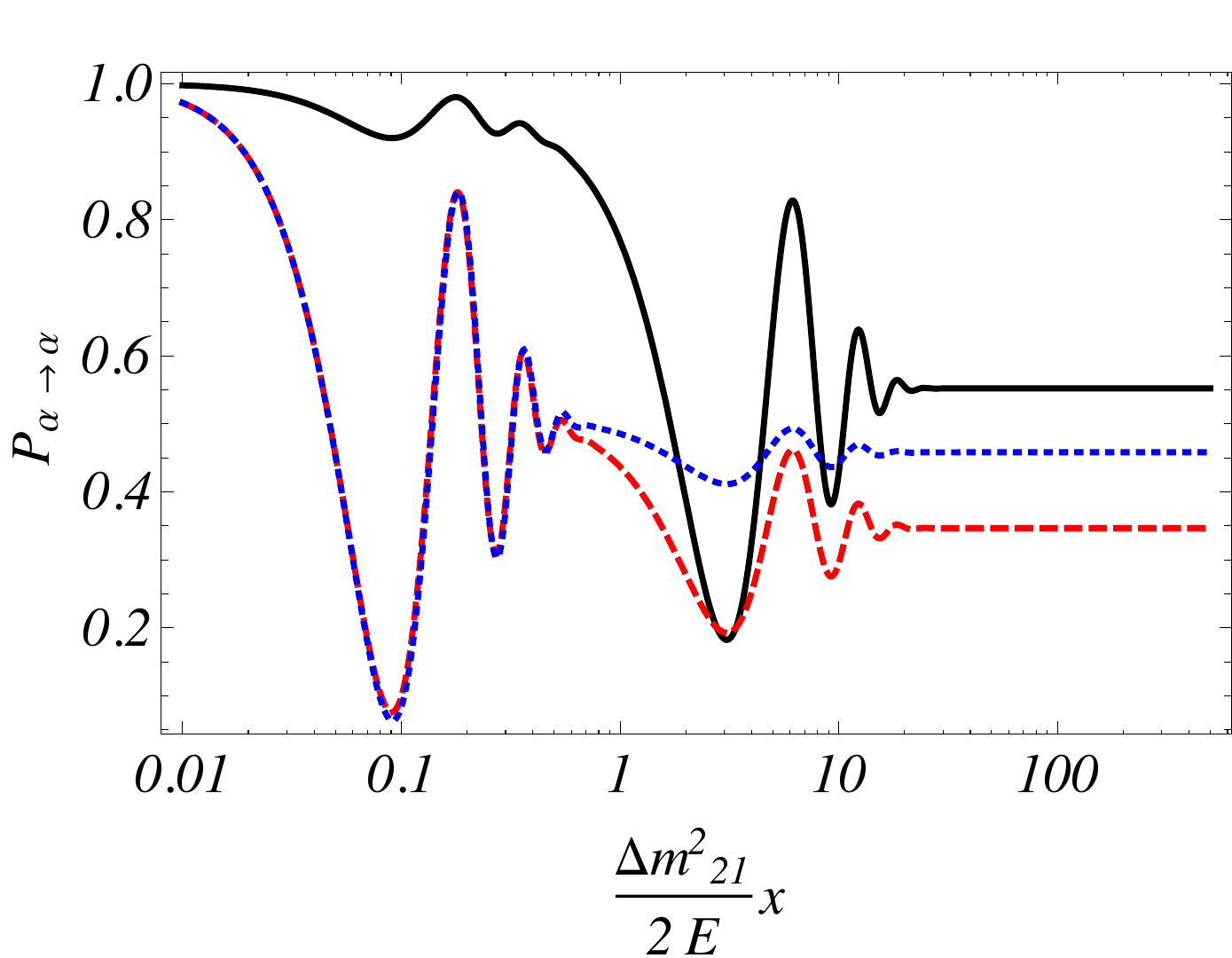}
\includegraphics[width = 7.0 cm]{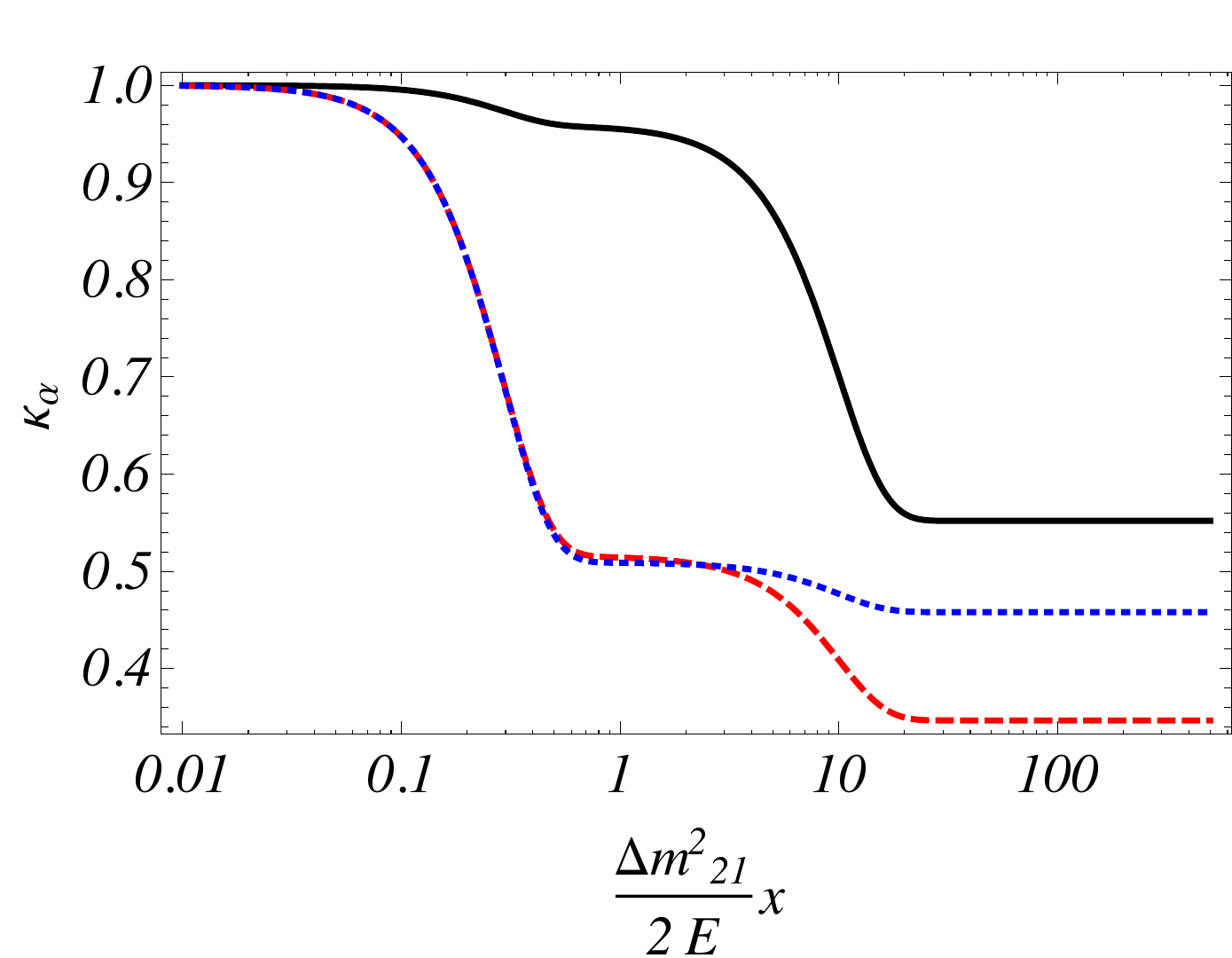}
\caption{(Color online) Survival probability (first plot) and quantum purity (second plot) as function of $\frac{\Delta m_{21}^2}{2 E} \, x$ for electronic ((black) solid lines), muonic ((red) dashed lines) and tauonic ((blue) dotted lines) for neutrino pure states at $x = 0$.
In the first plot, the damping behavior due to the decoherence caused by the mass-eigenstate wave packet mutual {\em slippage} drives the probability to constant values beyond the damping scale given by a coherence length $\sim \frac{2 \sqrt{2} E^2}{\sigma_p \Delta^2 m_{21}}$) for all flavors.
The second plot shows states evolving into statistical mixtures with the quantum purity prescribed by the phenomenological mixing parameters.}
\label{fig:01}
\end{figure}

Quantum flavor oscillations can be comprehended from the single particle quantum mechanical framework as a kind of three-level system problem. Starting with state vectors $\nu^{e}$, $\nu^{\mu}$ and $\nu^{\tau}$ related to electron, muon and taon neutrinos, respectively, one can identify the flavor state time-evolution as given by
\small
\begin{equation}
\left(\begin{array}{l}\nu^e_{(t)} \\ \nu^{\mu}_{(t)} \\ \nu^{\tau}_{(t)} \end{array}\right) =
U \,
D(t)
\,
\left(\begin{array}{l}\nu_1 \\ \nu_2 \\ \nu_3 \end{array}\right) =
U \,
D(t)
\, U^{\dagger} \left(\begin{array}{l} \nu^e_{(0)} \\ \nu^{\mu}_{(0)} \\ \nu^{\tau}_{(0)} \end{array}\right),
\label{101A}
\end{equation}
\normalsize
with $D(t) = Diag\left[ e^{-i\, E_{1} t},\,  e^{-i\, E_{2} t} ,\,  e^{- i\,E_{3} t} \right]$, $E_{k} = \sqrt{p^{2} + m^{2}_{k}}$, in natural units ($c=1$), with $k = 1,\, 2, \, 3$, and where $\nu_{1}$, $\nu_{2}$ and $\nu_{3}$ are the mass-(energy-)eigenstates and the mixing matrix, $U$, is given by
\small\begin{equation}
U =
\left(\begin{array}{ccc} c_{12} \, c_{13} & s_{12}\, c_{13} & s_{13} \, e^{-i \, \delta} \\
-s_{12}\, c_{23} - c_{12}\, s_{23}\, s_{13} e^{i \, \delta} & c_{12} \, c_{23} - s_{12} \, s_{23} \, s_{12} \, e^{i \, \delta} & s_{23} \, c_{13} \\ s_{12} \, s_{23} - c_{12}\, c_{23} \, s_{12} \, e^{i \, \delta} & - c_{12} \, c_{23} - s_{12} \, c_{23} \, s_{12} \, e^{i \, \delta} & c_{23} \, c_{13}\end{array}\right),
\label{eq00A}
\end{equation}
\normalsize
where the following shorthand notation has been adopted: $s_{kj} \equiv \sin \bb{\theta_{kj}}$, and $c_{kj} \equiv \cos \bb{\theta_{kj}}$. In this case one has three mixing angles,  $\theta_{12},\, \theta_{13}$, and $\theta_{23}$, and a unique free-phase parameter, $\delta$ \cite{01b01}.
The Hamiltonian in the mass-eigenstate basis is straightforwardly obtained from Eq.~(\ref{eq00A}) as $H \equiv Diag\{E_{1},\, E_{2}, E_{3}\}$, and after simple mathematical manipulations \cite{01b01,01b08,01b09}, the flavor oscillation probabilities can be computed as to give
\begin{eqnarray}
\mathcal{P}_{\alpha \rightarrow \alpha} \bb{t} = |\langle \nu^{\alpha}_{(0)}|\nu^{\alpha}_{(t)}\rangle|^{2}  = \displaystyle \sum_{k,j = 1}^3 \vert U_{\alpha k}\vert^2 \, \vert U_{\alpha j}\vert^2 \, \exp \left(- i (E_k - E_j) t \right),
\label{eq02CB}\\
\mathcal{P}_{\alpha\rightarrow \beta}\bb{t} = |\langle \nu^{\beta}_{(0)}|\nu^{\alpha}_{(t)}\rangle|^{2} = \displaystyle \sum_{k,j = 1}^3 \vert U_{\alpha k}\vert^2 \, \vert U_{\beta j}\vert^2 \, \exp \left(- i (E_k - E_j) t \right) ,
\label{eq02C}
\end{eqnarray}
namely the probabilities of $\alpha$-flavor states being created at time $t_{0}\sim 0$, and either being detected as $\alpha$-flavor states, or being converted into $\beta$-flavor states, both at a time $t > t_{0}$, with $\alpha$ and $\beta$ identified by flavors $e$, $\mu$, and $\tau$.
Localization effects that may change the above flavor oscillating pattern can be introduced, for instance, through a simplified one-dimensional wave packet prescription \cite{01b05,01b06} for each mass-eigenstate, $\vert \nu_k \rangle$, with
$\langle x \vert \nu_k (x, t) \rangle = \psi_k (x,t) \vert \nu_k \rangle$,
where
\begin{eqnarray}
\psi_k (x,t) &=& \frac{1}{(2 \pi)^{1/2}} \int \! \! dp \, \, \psi_k (p) \exp{\left(i [p\,x - E_k (p) t]\right)},
\end{eqnarray}
and, as a matter of convenience, the momentum distribution, $\psi_k (p)$, is a Gaussian function given by
\begin{equation}
\psi_k (p) = \frac{1}{(2 \pi \sigma_p ^2)^{1/4}} \exp{\left(- \frac{p^2 - p_k^2}{4 \sigma_p ^2}\right)}.
\end{equation}
The survival probabilities from Eq.~(\ref{eq02CB}) are thus converted into damped flavor oscillating relations as
\begin{equation}
\label{02}
P_{\alpha \rightarrow \alpha}(x,t) = \displaystyle \sum_{k,j = 1}^3 \vert U_{\alpha k} U_{\alpha j} \vert^2 \psi_k (x,t) \psi_j ^*(x,t),
\end{equation}
with $\alpha =$ $e$, $\mu$, and $\tau$.
Due to tiny (weak interaction) cross sections involved in the physics of neutrino detection, $P_{\alpha \rightarrow \alpha}(x,t)$ is measured by exposing a detector to continuous neutrino fluxes, which smears out any detector time resolution \cite{01b09}.
Meanwhile, it supports the statistics for a time-averaged expression resumed by
\begin{equation}
\label{04}
P_{\alpha \rightarrow \alpha} (x)=  \displaystyle \sum_{k,j = 1}^3 \vert U_{\alpha k} U_{\alpha j} \vert^2 f_{kj}(x),
\end{equation}
with $f_{kj} (x) = \int dt \, \, \psi_k(x,t) \psi_j ^*(x,t)$.
Assuming that mass-eigenstate wave packets propagate in a relativistic regime, with $m_k << p$, such that \cite{01b01}
\begin{equation}
E_k (p) = \sqrt{p^2 + m_k^2} \simeq\sqrt{p_k ^2 + m_k ^2} + v_k (p - p_k) = E_k + v_k (p - p_k),\end{equation}
with $E_k \simeq E$, $p_k \simeq E -  m_k ^2/2E$, and $v_k = p_k/ E\simeq  1 - m_k ^2/2E^{2}$, one can approximate $f_{kj} (x)$ by
\begin{equation}
f_{kj} (x) \approx \exp\left[ -i \frac{\Delta m_{kj} ^2}{2 E} x - \left( \frac{\sigma_p \Delta m_{kj} ^2 x }{2 \sqrt{2} E^2}\right)^2 \right],
\end{equation}
where $\Delta m_{kj} ^2 = m_k ^2 - m_j^2$. Substituting the above result into Eq.~(\ref{04}) and introducing numerical values to the phenomenological parameters as to have $\sin^2 \bb{\theta_{12}} = 0.306$, $\sin^2\bb{\theta_{13}} = 0.021$,  and $\sin^2 \bb{ \theta_{23} } = 0.42$ for the mixing angles, and $\Delta m_{21} ^2 = 2.35 \times 10^{-3} \, eV^2$, and $\Delta m^2 _{32} = 7.58 \times 10^{-5} \, eV^2$ for the squared mass differences, one obtains the survival probabilities as depicted in the first plot of Fig.~\ref{fig:01}.
The oscillation probabilities exhibit a damping effect due to the mass-eigenstate wave packet decoherence, a typical spatial delocalization effect featured by a coherence length $\sim \frac{2 \sqrt{2} E^2}{\sigma_p \Delta^2 m_{21}}$.

To generalize the above results to a composite quantum system framework, one can introduce a localized density matrix operator, $\rho_\alpha (x)$, and extend the time-averaged procedure \cite{01b09} as to have
\begin{equation}
\label{05}
\rho_\alpha(x) = \int \! \! dt \, \, \rho_\alpha (x,t) = \displaystyle \sum_{k,j = 1}^3 U_{\alpha  k} \, U_{\alpha  j}^* \, f_{k j} (x) \, \vert \nu_k \rangle \langle \nu_j \vert,
\end{equation}
such that flavor projection operators, $M^{\alpha}$, can be identified as
\begin{equation}
M^{\alpha} = \displaystyle \sum_{k,j = 1}^3 \, U_{\alpha  k}  U^*_{\alpha j} \, \vert \nu_k \rangle \langle \nu_j \vert,
\end{equation}
through which one sets a simplified expression for the survival probabilities \cite{01b08},
\begin{equation}
P_{\alpha \rightarrow \alpha} (x) = Tr[M_\alpha \, \rho_\alpha (x)].
\end{equation}

To quantify the degree of statistical mixture for a system initially created ($t = 0$) as a pure flavor state, $\alpha$, the quantum purity is defined as
\begin{equation}
\label{06}
\kappa_\alpha (x) = Tr[\rho_\alpha^2 (x)] = 1 - \displaystyle \sum_{k = 1} ^{2} \sum_{j= k + 1} ^3 2 \vert U_{\alpha  k} U_{\alpha  j} \vert^2 [1 - \vert f_{kj} (x)\vert ^2].
\end{equation}
The quantum purity for three-flavor quantum systems, with the phenomenological parameters above introduced, is depicted in the second plot of Fig.~\ref{fig:01}.
As one can notice, in spite of not exhibiting an oscillation pattern, the quantum purity, $\kappa_\alpha (x)$, is suppressed as it evolves along the localization scale.
The function $\vert f_{kj} (x) \vert$ follows a decreasing profile which asymptotically vanishes for $x$ above a certain value that sets the damping scale.
The decreasing behavior of the quantum purity hence describes the configuration of mixed states with the level of mixing determined driven by mass differences and mixing angles, and by a localization scale.

\paragraph{Mass eigenstates, {\em qubits} and entanglement quantifiers.}

Entanglement quantifiers correspond to quantum correlations that measure the separability of composite quantum systems.
A state that describes a system composed by $n$ subsystems in a Hilbert space $\mathcal{H} = \bigotimes_{k = 1} ^n \mathcal{H}_k$ is separable if and only if it can be written as
\begin{equation}
\vert \psi \rangle
= \vert \psi_1 \rangle \otimes \vert \psi_2 \rangle \otimes ... \otimes \vert \psi_n \rangle
=\bigotimes_{k = 1} ^n \vert \psi_k \rangle,
\end{equation}
with $\vert \psi_k \rangle \in \mathcal{H}_k$ \cite{01b11}.
A composite state described by a density operator $\rho$ is separable if it can be decomposed into $\rho = \displaystyle \sum_{k = 1}^s p_k \bigotimes_{j = 1} ^n \rho_{j} ^k$,
with $\rho_{j} ^k$ acting on $\mathcal{H}_j$ \cite{01b15}.
If a state is not separable then it is entangled \cite{01b11}.

In the neutrino scenario, if one notices that mass-eigenstates are orthonormal, $\langle \nu_k \vert \nu_j \rangle = \delta_{kj}$, in a $3$-dim Hilbert space spanned by $\{\vert \nu_k \rangle \}$, one can interpret the label $k$ as the index for the quantum mode \cite{01b10} that relates mass-eigenstates and three-{\em qubit} states.
One thus has
$\vert \nu_k \rangle = \vert \delta_{k1} , \delta_{k2} , \delta_{k3} \rangle$,
where {\em qubits} are interpreted as the number of occupation in the respective mass mode.
Flavor-eigenstates are then described as a superposition of {\em qubit} states,
\begin{equation}
\vert \nu_\alpha \rangle = \displaystyle \sum_{k = 1} ^3 U_{\alpha  k}\vert \delta_{k1}, \delta_{k2}, \delta_{k3} \rangle =  U_{\alpha  1} \vert 1 0 0 \rangle + U_{\alpha 2} \vert 010 \rangle + U_{\alpha   3} \vert 001 \rangle ,
\end{equation}
which, in the composite quantum system framework, support the corresponding density matrix written as
\begin{equation}
\label{13}
\rho_\alpha (x) = \displaystyle \sum_{k,j = 1} ^3 U_{\alpha  k} \, U^*_{\alpha  j} \, f_{kj} (x) \vert \delta_{k1},  \delta_{k2}, \delta_{k3} \rangle \langle \delta_{j1},  \delta_{j2} , \delta_{j3} \vert,
\end{equation}
which turns into a $W$-like entangled state for $x = 0$ \cite{01b16}.
As the system evolves as a function of the variable $x$, the entanglement between the mass modes changes as $x$ increases.

To identify any correlation between the damping oscillation and the quantum entanglement for pure states, for instance, one can consider the entropy of entanglement \cite{01b11} as a suitable quantifier of the entanglement between two subsystems.
A pure state $\rho = \vert \psi \rangle \langle \psi \vert$ only admits a unique Schmidt decomposition \cite{01b11} that is given in terms of base vectors of each Hilbert space related to subsystems $A$ and $B$, and of the Schmidt coefficients of the pure state decomposition.
If it exhibits some level of entanglement between the two subsystems $A$ and $B$, then the reduced representation of any of these subsystems, either $\rho_A = Tr_{B} [\rho]$ for the subsystem $A$, or $\rho_B = Tr_{A}[\rho]$ for the subsystem $B$, results into mixed states.
One infers that a pure state is entangled if one of its reduced systems is mixed.
The entropy of entanglement between two subsystems that compose a pure state, $\rho$, is therefore defined by
$S[\rho] = S[\rho_A] = S[\rho_B] = - \sum \lambda_k \log{\lambda_k}$,
where $S[\rho]$ is the von Neumann (vN) entropy \cite{01b11} of the corresponding density matrix which is computed in terms of the $\rho$ eigenvalues, $\lambda_k$.
Pure states in the reduced representation, $\rho_{A,B}$, set the entropy of entanglement equals to zero.
It implies that the state $\rho$ is separable.
Likewise, reduced mixed states lead to non-vanishing values to the vN entropy, which implies into an entangled state, $\rho$.

Mixed quantum states do not have a unique decomposition into pure states \cite{01b17} such that the entropy of entanglement is no longer useful in this case.
Hence the $E_{F}$ \cite{01b18} and the $E_{\mathcal{N}}$ \cite{01b19} are indeed the most suitable entanglement quantifiers for mixed states.

For a mixed state, $\rho = \displaystyle \sum_{k} p_k \vert \psi_k \rangle \langle \psi_k \vert$, for which all possible pure state decompositions with respective probabilities, $p_k$, have been considered, its corresponding $E_{F}$ is defined as the average entanglement of the pure states of the decomposition, minimized over all decompositions of $\rho$ \cite{01b18}, i. e. $E_{F} [\rho] = \mbox{inf} \,  \displaystyle \sum_{k} p_k \, S( \, \vert \psi_k \rangle \langle \psi_k \vert\, )$.
For a pair of {\em qubits}, the above specified minimum value is given by \cite{01b20}
\begin{equation}
\label{15}
E_{F} [\rho ] = \mathcal{E}[C(\rho)] = h \left( \frac{1 + \sqrt{1 - C^2}}{2}\right),
\end{equation}
where $h(x) = - x \log_2 x - (1 - x)\log_2(1 - x)$, and
$C \equiv C[\rho] = \max\{ 0,  \lambda_1 - \lambda_2 - \lambda_3 - \lambda_4\}$ is the quantum concurrence \cite{01b20}, with $\lambda_1 \ge \lambda_2 \ge \lambda_3 \ge \lambda_4$ corresponding to the square roots of the eigenvalues of $\rho \, (\sigma_y \otimes \sigma_y) \, \rho^* \, (\sigma_y \otimes \sigma_y)$.
The quantum concurrence of a bipartition $\rho_\alpha ^{\{i,j\}} (x) = Tr_{k \neq\{i,j\}} [\rho_\alpha(x)]$ of (\ref{13}) is thus given by
\begin{equation}
C_{\alpha}^{\{ i,j \}} = 2 \vert U_{\alpha  i}\, U_{\alpha j} \vert \, \vert f_{ij} (x) \vert,
\end{equation}
and the $E_{F}$ between two modes, $E_{F, \alpha} ^{\{i,j\}}$, can be straightforwardly obtained from (\ref{15}).
The averaged $E_{F}$ is thus defined by
\begin{equation}
\label{16}
\overline{E}_{F , \alpha} (x) = \frac{1}{3}[ E_{F , \alpha} ^{\{1,2\}} (x) + E_{F , \alpha} ^{\{1,3\}} (x) + E_{F , \alpha} ^{\{2,3\}} (x)].
\end{equation}

Likewise, for a bipartite system $\rho$, one defines the negativity, $\mathcal{N}(\rho)$, as the absolute value of the sum of negative eigenvalues of $\rho^{T}$, which is obtained through the partial transposition of $\rho$ with respect to one mode such that, given an arbitrary orthonormal basis $\vert i , j \rangle$, the matrix elements of $\rho^T$ are obtained through $\langle i , j \vert \rho^T \vert i^\prime , j^\prime \rangle = \langle i^\prime , j \vert \rho \vert i , j^\prime \rangle $.
The $E_{\mathcal{N}}$ is thus defined by \cite{01b19}
\begin{equation}
E_{\mathcal{N}} = \log_2[1 + 2 \, \mathcal{N}(\rho)].
\end{equation}

Considering the bipartition of the tripartite system (\ref{13}) into two subsystems: $\{ i \}$ and $\{j,k \}$, the $E_{\mathcal{N}}$ associated to the fixed bipartition $\{i;j,k\}$ shall be given by
\begin{equation}
E_{\mathcal{N},\alpha} ^{\{ i; j, k\}} (x) = \log_2[1 + 2\vert U_{\alpha  i} \vert \sqrt{\vert U_{\alpha  k} \vert^2 \vert f_{ki} (x) \vert^2 + \vert U_{\alpha  j} \vert^2 \vert f_{ij} (x) \vert^2 } ],
\end{equation}
and the corresponding averaged $E_{\mathcal{N}}$ associated to a flavor-eigenstate is identified by
\begin{equation}
\label{17}
\overline{E}_{\mathcal{N},\alpha} (x) = \frac{1}{3}[E_{\mathcal{N},\alpha} ^{\{ 1; 2, 3\}} (x) + E_{\mathcal{N},\alpha} ^{\{ 2; 1, 3 \}} (x) + E_{\mathcal{N},\alpha} ^{\{ 3; 1, 2\}} (x)].
\end{equation}

\begin{figure}
    \includegraphics[width = 8 cm]{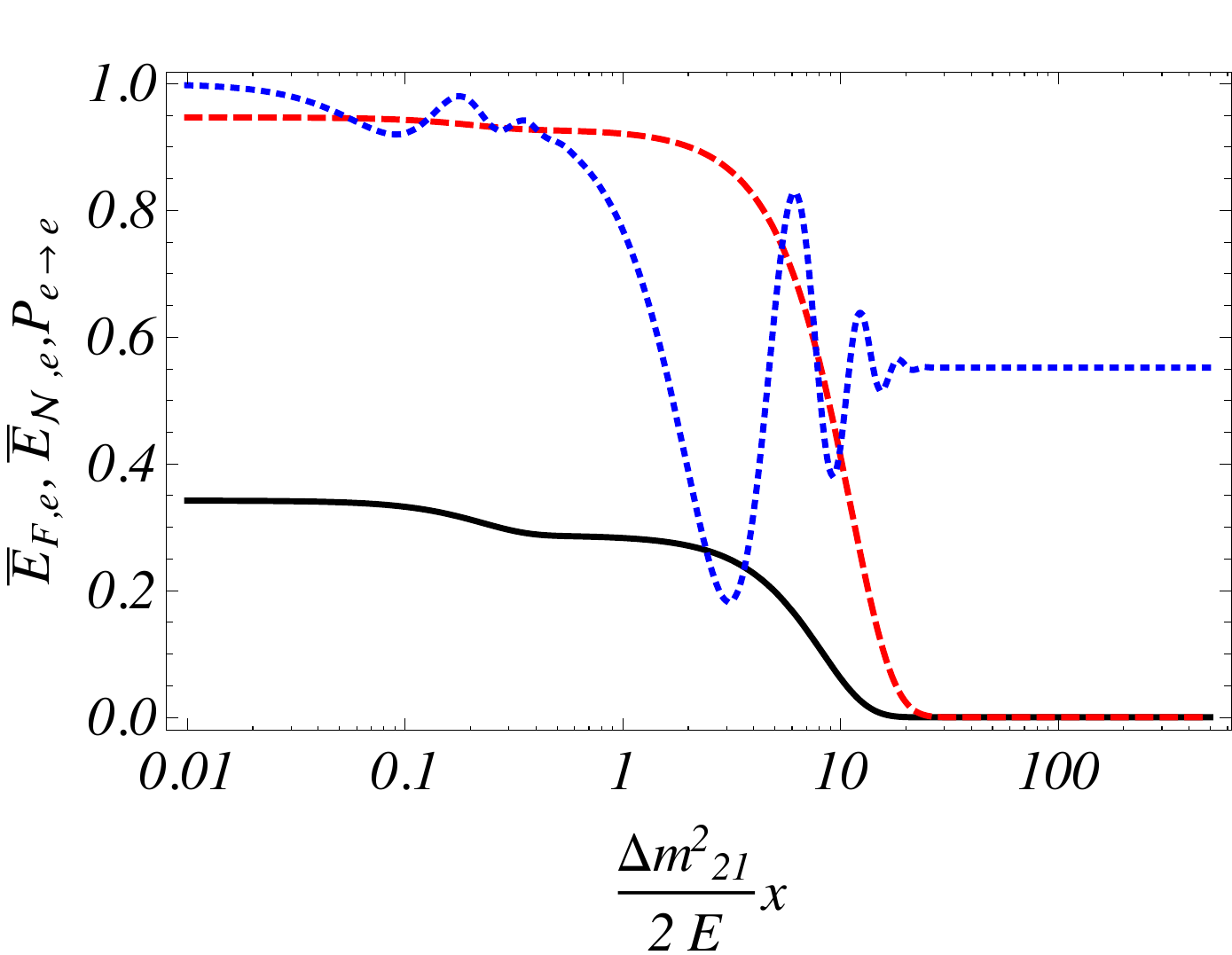}
    \includegraphics[width = 8 cm]{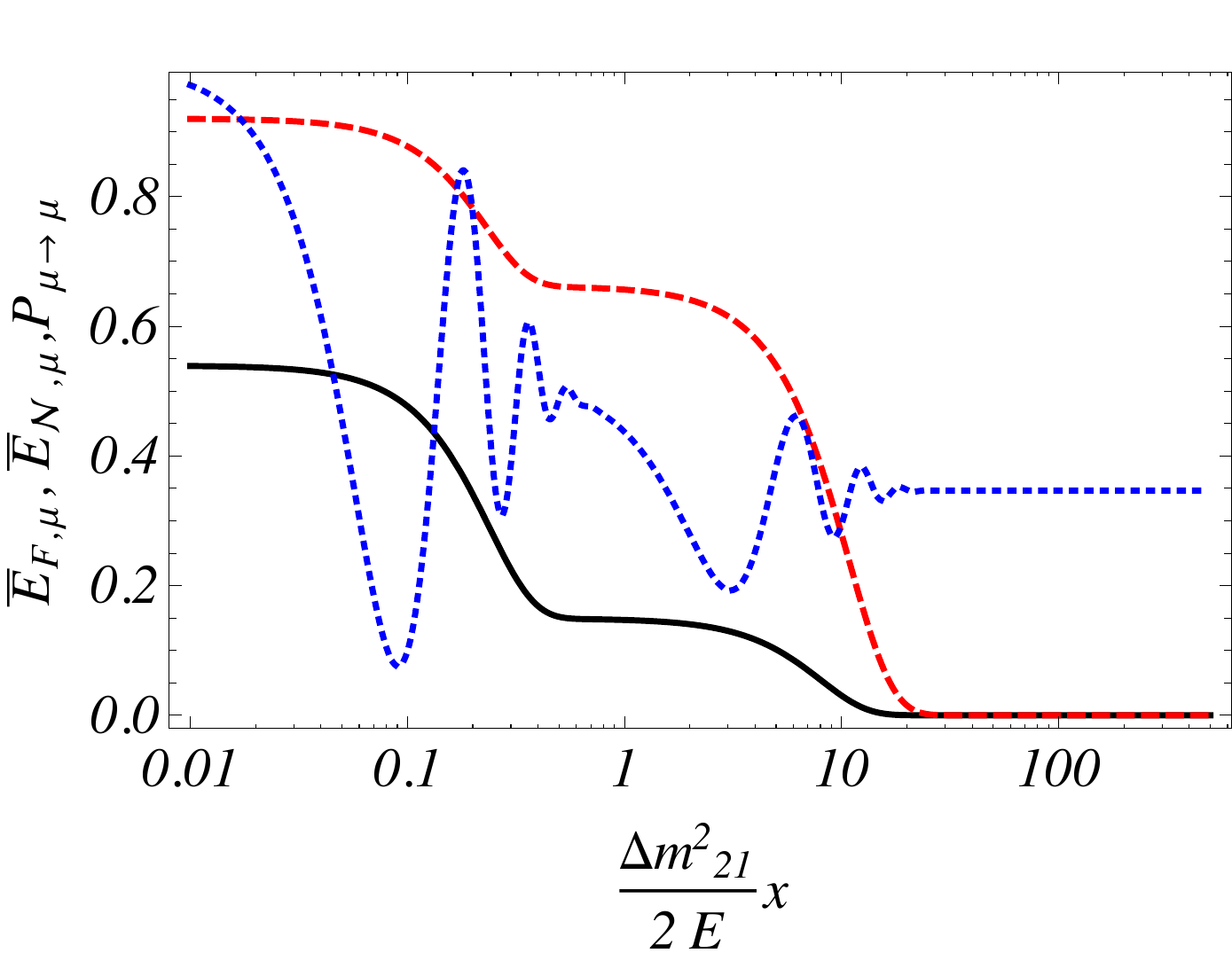}
    \includegraphics[width = 8 cm]{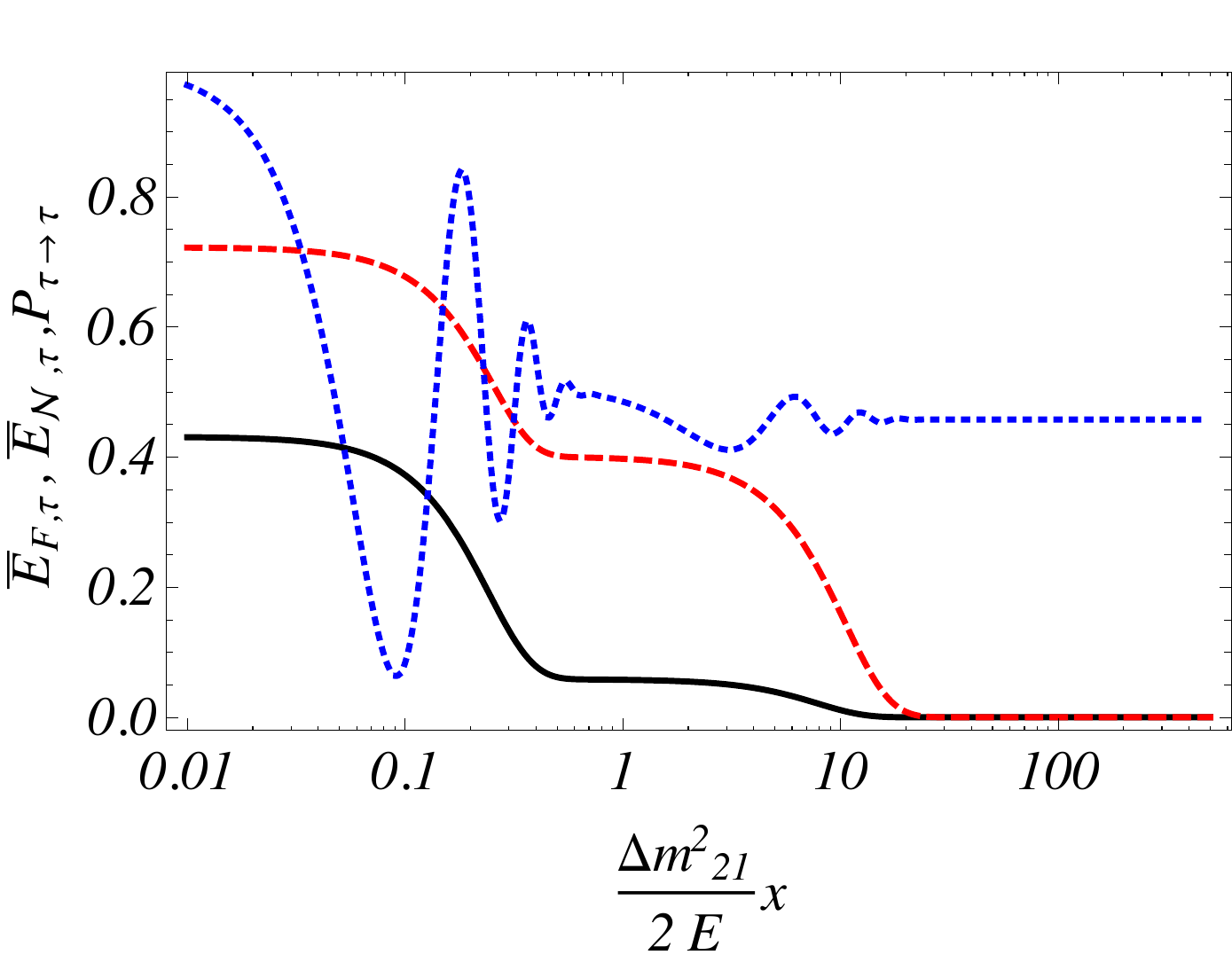}
\caption{ \label{fig:03} (Color online) Averaged $E_{F}$ ((black) solid lines) and $E_{\mathcal{N}}$ ((red) dashed lines) as function of $\frac{\Delta m_{21} ^2}{2 \, E} \, x $. The plots are for eletronic (left) , muonic (middle) and tauonic (right) neutrino flavors in correspondence with the survival probability ((blue) dotted lines) for $\sigma_p / E = 1$.
Notice that the coherence scale and the damping scale of survival probabilities approach each other.}
\end{figure}

The above introduced averaged quantities for $E_{F}$ and $E_{\mathcal{N}}$ as given by Eqs.~(\ref{16}) and (\ref{17}), are depicted in Fig.~\ref{fig:03} and confronted with damped flavor oscillations described by the corresponding survival probabilities.

Decoherence and loss of entanglement between subsystems are driven by the dependence on the localization parameter, $f_{kj} (x)$, as a function of $x$. The decoherence scale, i.e. the scale above which entanglement is negligible, is comparable with the damping scale. A qualitative coincidence is expected as the two phenomena are permeated by the same delocalization effect: the gradual separation of mass-eigenstate wave packets.

\paragraph{Maximal correlation between flavor entanglement and damped oscillations.}

Localization and decoherence effects brought up by the wave packet picture suggest the existence of tiny connection between flavor entanglement and oscillation damping in the scenario of three neutrino mixing.
The previously obtained results indicate that a global derivative of the survival probabilities fix a set of damping scale end-points for flavor oscillating probabilities. The localization effects introduce a coherence length that resumes the gradual separation of mass-eigenstates. The entanglement quantifiers depend strictly on $\vert f_{kj} (x) \vert$ instead of $f_{kj}$, as it appears in the analytical expressions for oscillation damping of survival probabilities.
It creates an unambiguous limitation in connecting both results.

Otherwise, giving an entropic interpretation to the survival probability, which can be parameterized by a Boltzmann density of states such that $ \Omega_{\alpha}(x) /\Omega \rightarrow P_{\alpha \rightarrow \alpha}(x)$, one can identify an \textit{ad hoc} statistical correlation between survival probabilities, $P_{\alpha \rightarrow \alpha}(x)$, and an equivalent Boltzmann entropy, $\Delta \mathcal{S}_\alpha(x) = \mathcal{S}(x) - \mathcal{S}_{\alpha}(x) = k_B \ln(\Omega_{\alpha}(x) /\Omega)$.
It shall be quantified by a deviation parameter, $\chi_{\mathcal{S}}$, which is set equal to
\begin{equation}
\chi_{\mathcal{S}, \alpha}(x) \sim \frac{d}{dx}\left[ P_{\alpha \rightarrow \alpha}(x) - \exp(\Delta \mathcal{S}_\alpha(x))\right].
\label{chi}
\end{equation}
$\chi_{\mathcal{S}}$ essentially depends on localization parameters, $x$ and $\sigma_p / E$, for which the derivative with respect to $x$ suppresses any scale shift effect.
Eq.~(\ref{chi}) may be used to compute the statistical correlation between survival probabilities and entanglement quantifiers, as $\Delta \mathcal{S}$ can be replaced by either $E_{F}$ or $E_{\mathcal{N}}$ into Eq.~(\ref{chi}).

The mean square deviation of $\chi$ over $x$ is defined by
\begin{equation}
\Delta \chi = \sqrt{\langle \chi_{\mathcal{S},\alpha}^2 \rangle - \langle \chi_{\mathcal{S}, \alpha} \rangle ^2},
\end{equation}
with $\langle \chi_{\mathcal{S}, \alpha} \rangle = \int dx \mbox{ } \chi_{\mathcal{S},\alpha}$ and $\langle \chi_{\mathcal{S},\alpha}^2 \rangle = \int dx \mbox{ } \chi_{\mathcal{S},\alpha} ^2$.
It can be used to estimate the effects of wave packet localization on the correlation between damping and entanglement.

\begin{figure}
    \includegraphics[width = 14.0 cm]{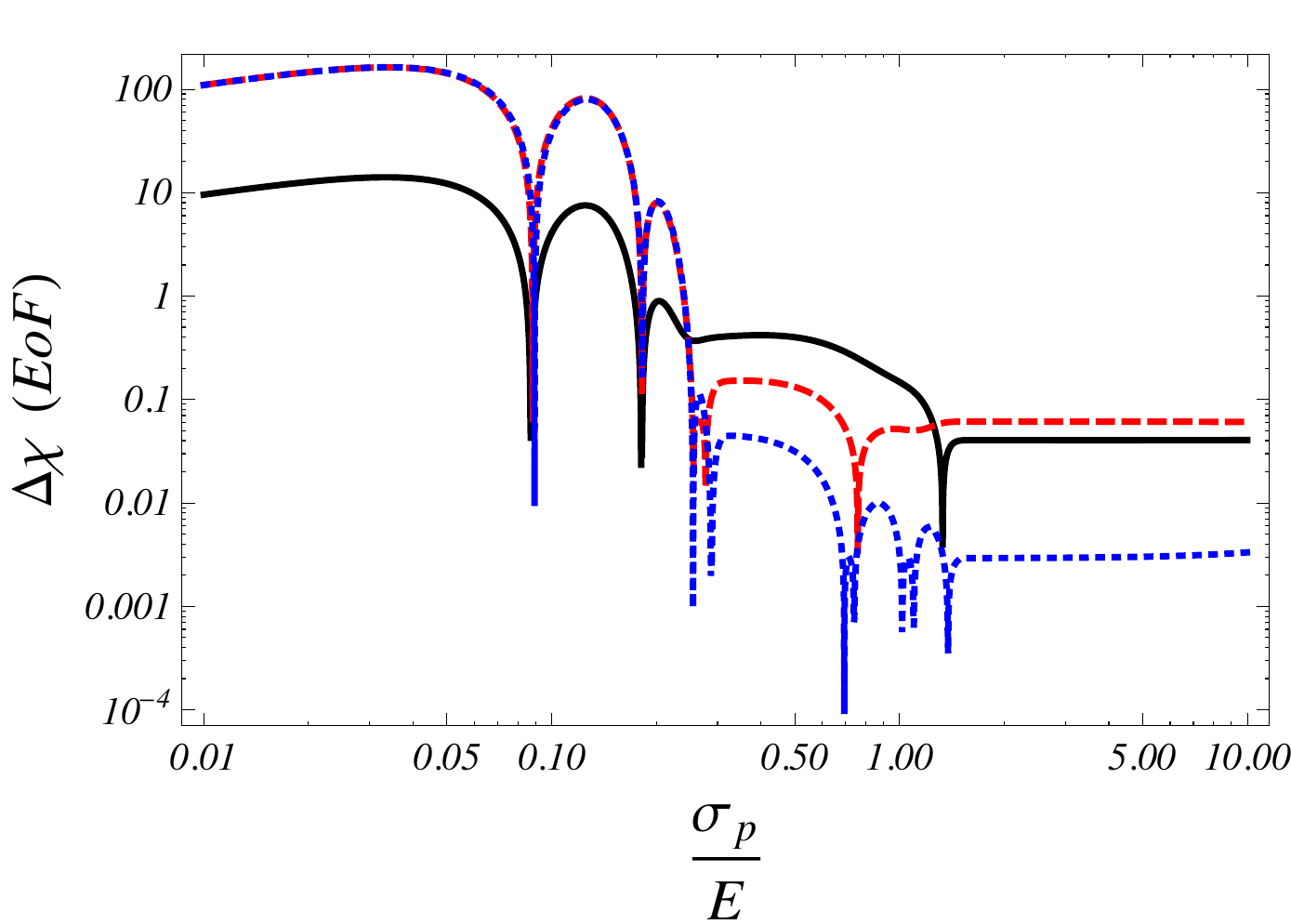}
    \includegraphics[width = 14.0 cm]{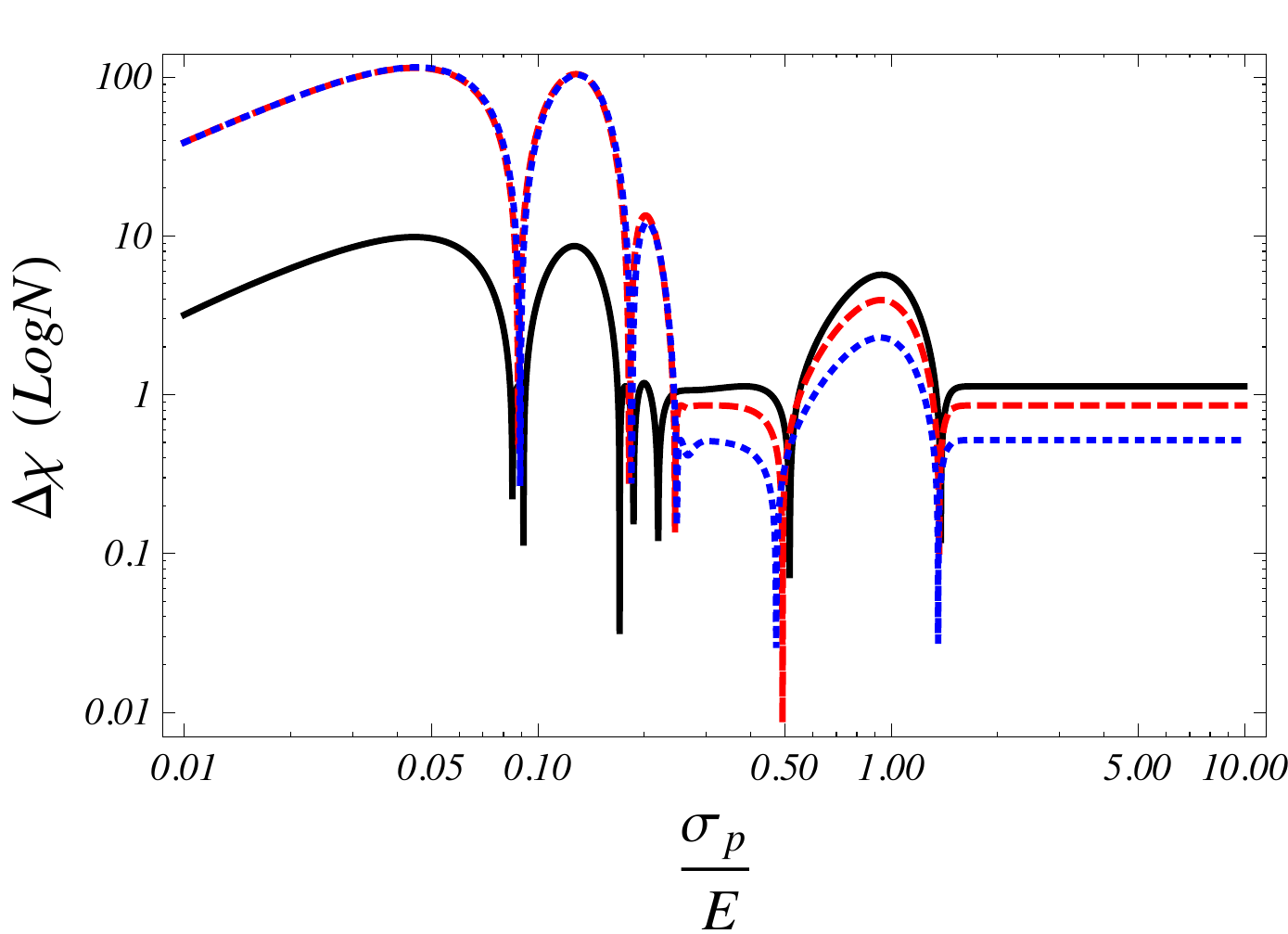}
\caption{\label{fig:05} (Color online) Mean square deviation $\Delta \chi$ for the entanglement of formation (first plot) and logarithmic negativity (second plot) for electronic ((black) solid line), muonic ((red) dashed line) and tauonic ((blue) dotted line) neutrinos as function of $\sigma_p/E$.}
\end{figure}

Fig.~\ref{fig:05} shows that the statistical correlation between $P_{\alpha \rightarrow \alpha}$ and $\exp(\Delta \mathcal{S}_\alpha)$, as stated by $\Delta \chi$, is highly affected by the localization parameter $\sigma_p/E$.
The correlation quantified by $\Delta \chi$ can be maximized for some specific choices of  $\sigma_p/E$.
As presumed by the phenomenology, the results are constrained by neutrino mass and mixing angles and production/detection localization parameters.
The maximal correlation between flavor entanglement and oscillation damping can be identified by the values of $\sigma_p/E$ that correspond to the set of minimal points of the plotted curves, for which $\Delta \chi$ vanishes.

\paragraph{Conclusions}

Quantum entanglement and oscillation damping in the context of flavor quantum transition are show to have the same origin owing to localization and decoherence effects brought up by an external wave packet picture.
Describing flavor-eigenstates as composite quantum systems has provided theoretical tools for obtaining a decoherence scale driven by the loss of quantum entanglement supported by a three-{\em qubit} partition where each {\em qubit} has been identified as a neutrino mass mode.
A statistical correlation between the oscillation damping scale and the mass mode separability scale has been quantified in terms of an effective localization parameter, $\sigma_p/E$.
The maximal correlation depicted in Fig.~\ref{fig:05} shows that the correspondence
between quantum entanglement and oscillation damping due to localization effects is highly dependent on the localization scale, for which utmost correlations involving the entanglement of formation and the logarithmic negativity have been identified.

\paragraph{Acknowledgments}
 This work was supported by the Brazilian Agency CNPq (grant 300809/2013-1, grant 139843/2012-4 and grant 307789/2012-8), and by the National Institute for Science and Technology of Quantum Information (INCT-IQ).

\pagebreak
\newpage

\end{document}